\documentclass[twocolumn,aps,prl,showpacs]{revtex4}

\usepackage{latexsym,amssymb,float}
\usepackage{setspace}
\usepackage{graphicx}
\usepackage{epsfig}
\usepackage{amsmath}
\usepackage{bm}

\begin{document}

\title{Hidden Order Transition in URu$_2$Si$_2$ and the Emergence of a Coherent Kondo Lattice}
\author{Ting Yuan, Jeremy Figgins and Dirk K. Morr}
\affiliation{Department of Physics, University of Illinois at
Chicago, Chicago, IL 60607, USA}
\date{\today}

\begin{abstract}

Using a large-$N$ approach, we demonstrate that the differential
conductance and quasi-particle interference pattern measured in
recent scanning tunneling spectroscopy experiments (A.R. Schmidt
{\it et al.} Nature {\bf 465}, 570 (2010); P. Aynajian {\it et al.},
PNAS {\bf 107}, 10383 (2010)) in URu$_2$Si$_2$ are consistent with
the emergence of a coherent Kondo lattice below its hidden order
transition (HOT). Its formation is driven by a significant increase
in the quasi-particle lifetime, which could arise from the emergence
of a yet unknown order parameter at the HOT.

\end{abstract}


\maketitle

Heavy-fermion materials exhibit a plethora of exciting phenomena
\cite{Exp} which are believed to arise from the competition between
Kondo screening \cite{Kondo64} and antiferromagnetic ordering
\cite{Don77}. One of the most puzzling phenomena arises in the
heavy-fermion compound URu$_2$Si$_2$ which exhibits an onset of
Kondo screening around $T \approx 55$K \cite{Pal85,Map86}, and
undergoes a second order phase transition at $T_0 =17.5$K
\cite{Pal85,Map86,HOT} into a state with a still unknown (hidden)
order parameter. Currently, an intense debate focuses on the nature
of this {\it hidden order transition} (HOT) and its microscopic
origin \cite{Map86,HOT_Th}. Important new insight into this question
has recently been provided by groundbreaking scanning tunneling
spectroscopy (STS) experiments \cite{Schm09,Ayn10}. Above the HOT,
the differential conductance, $dI/dV$, exhibits a characteristic
Fano lineshape \cite{Fano61}. In contrast, below $T_0$, a soft gap
opens up in $dI/dV$ \cite{Schm09,Ayn10} and a quasi-particle
interference (QPI) analysis reveals a band structure similar to that
expected in the (heavy Fermi liquid) phase of a screened Kondo
lattice \cite{Schm09}. Whether the observed changes in $dI/dV$ below
the HOT are consistent with the observed QPI pattern and with the
emergence of a coherent Kondo lattice, is an important question
whose answer will provide crucial insight into the nature of the
hidden order transition.

In this Letter, we address this question and demonstrate that the
experimentally observed $dI/dV$ \cite{Schm09,Ayn10} and QPI pattern
\cite{Schm09} below the HOT are consistent with the emergence of a
coherent Kondo lattice (CKL) and its electronic band structure. In
particular, $dI/dV$ exhibits characteristic signatures of the Kondo
lattice band structure \cite{Ayn10}, such as an asymmetric gap, and
a peak inside the gap which arises from the van Hove singularity of
the heavy $f$-electron band. In addition, the temperature evolution
of $dI/dV$ \cite{Schm09,Ayn10} suggests that the formation of the
CKL below the HOT is primarily driven by a significant increase in
the coherence, i.e., lifetime, of the heavy quasi-particles. Since
the creation of a CKL is not expected to be the primary source of
the observed second order phase transition at $T_0$, we suggest that
the increased quasiparticle coherence is a result of the yet unknown
order parameter that emerges at the HOT.

The starting point for our study is the Kondo-Heisenberg Hamiltonian
\cite{Si01,chi,Col07}
\begin{eqnarray}
{\cal H} &=& \sum_{{\bf k},\sigma} \varepsilon_{\bf k}
c^\dagger_{{\bf k},\sigma} c_{{\bf k},\sigma} + J {\sum_{{\bf
r},\alpha,\beta}} {\bf S}^{K}_{\bf r} \cdot c^\dagger_{\bf {\bf r},
\alpha}{\bm \sigma}_{\alpha \beta}c_{{\bf {\bf r}}, \beta} \nonumber \\
& & + {\sum_{{\bf r,r'}}} I_{{\bf r,r'}} {\bf S}^{K}_{\bf r} \cdot
{\bf S}^{K}_{\bf r'} \label{eq:1} \ .
\end{eqnarray}
We use a hole-like two-dimensional (2D) conduction ($c$-electron)
band dispersion $\varepsilon_{\bf k}=2t(\cos k_x + \cos k_y )-\mu$
with nearest-neighbor hopping $t$ and chemical potential $\mu$ to
reproduce the 2D QPI dispersion \cite{Schm09}.  $c^\dagger_{{\bf
k},\alpha} (c_{{\bf k},\alpha})$ creates (annihilates) a
$c$-electron with spin $\alpha$ and momentum ${\bf k}$. $J>0$ is the
Kondo coupling, ${\bf S}^{K}_{\bf r}$ is the $S=1/2$ spin operator
of a magnetic atom at site ${\bf r}$ and ${\bm \sigma}$ are the
Pauli matrices. $I_{{\bf r,r'}}$ is the antiferromagnetic coupling
between magnetic atoms. In the {\em large-N} approach
\cite{largeN,Hew97}, one represents ${\bf S}^{K}_{\bf r}$ by
pseudo-fermion operators, $f^\dagger,f$, and decouples the
Hamiltonian via the mean fields \cite{Igl97,Kaul07,Fig10b}
\begin{equation}
s({\bf r}) = \frac{J}{2}\sum_{\alpha} \langle f^\dagger_{{\bf
r},\alpha} c_{{\bf r},\alpha} \rangle ; \quad \chi({{\bf r,r'}}) =
\frac{I_{\bf r,r'}}{2}\sum_{\alpha} \langle f^\dagger_{{\bf
r},\alpha} f_{{\bf r'},\alpha} \rangle. \label{eq:SC1}
\end{equation}
A non-zero hybridization $s({\bf r})$ between the $c$- and
$f$-electron states describes the screening of a magnetic moment,
and the bond variable $\chi({\bf r},{\bf r}')$ represents the
antiferromagnetic (spin-liquid) correlations \cite{chi} between
nearest-neighbor moments. For a translationally invariant system,
$s({\bf r})=s$, $\chi({\bf r},{\bf r}')=\chi_0$ and $\chi_1$ for
nearest and next-nearest-neighbor sites, respectively. Adding the
term $\sum_{{\bf r},\alpha} \varepsilon_f f^\dagger_{{\bf r},\alpha}
f_{{\bf r},\alpha}$ to the Hamiltonian \cite{Igl97,Kaul07,Fig10b}
allows one to fix the $f$-electron occupancy, $\langle {\hat n}_f
\rangle$, by adjusting the on-site energy $\varepsilon_f$. To solve
the self-consistency equations, Eq.(\ref{eq:SC1}), for finite
lifetimes of the $f$- and $c$-electron states, we rewrite them in
the form
\begin{eqnarray}
s({\bf r}) &=& -\frac{J}{\pi} \int_{-\infty}^{\infty} d\omega \
n_F(\omega) \ {\rm
Im} G_{fc}({\bf r},{\bf r},\omega)  \ ; \nonumber \\
\chi({{\bf r,r'}}) &=& -\frac{I_{\bf r,r'}}{\pi}
\int_{-\infty}^{\infty} d\omega \ n_F(\omega) \ {\rm Im} G_{ff}({\bf
r},{\bf r^\prime},\omega)  \ , \label{eq:SC2}
\end{eqnarray}
where $G_{\gamma \zeta}({\bf r}^\prime, {\bf r}, \tau)=-\langle
T_\tau \gamma_{{\bf r}^\prime}(\tau) \zeta^\dagger_{\bf r}(0)
\rangle$ ($\gamma,\zeta=c,f$, spin indices are omitted) are the full
Green's function describing the hybridization of the $c$- and
$f$-electron bands, with
\begin{eqnarray}
G_{ff}({\bf k}, \omega) & = &  \left[(G_{ff}^0({\bf k}, \omega))^{-1} - s^2 G_{cc}^0({\bf k}, \omega) \right] ^{-1} \ ; \nonumber \\
G_{cc}({\bf k}, \omega) & = & \left[(G_{cc}^0({\bf k}, \omega))^{-1} - s^2 G_{ff}^0({\bf k}, \omega) \right] ^{-1} \ ; \nonumber \\
G_{cf}({\bf k}, \omega) & = & G_{cc}^0({\bf k}, \omega) s
G_{ff}({\bf k}, \omega) \ , \label{eq:GF}
\end{eqnarray}
where $G_{ff}^0 = (\omega + i \Gamma_f -\chi_{\bf k})^{-1}, G_{cc}^0
= (\omega + i \Gamma_c -\varepsilon_{\bf k})^{-1}$, and
$\Gamma^{-1}_c$ and $\Gamma^{-1}_f$ are the lifetimes of the $c$-
and $f$-electron states, respectively. For $\Gamma_c=\Gamma_f=0^+$,
the poles of the above Green's functions yield two energy bands
\begin{equation}
E_{\bf k}^{\pm}=\frac{\varepsilon_{\bf k} + \chi_{\bf k}}{2} \pm
\sqrt{\left( \frac{\varepsilon_{\bf k} - \chi_{\bf k}}{2}\right)^2 +
s^2}
\end{equation}
with $\chi_{\bf k} = -2\chi_0(\cos k_x+\cos k_y)-4\chi_1\cos k_x
\cos k_y + \varepsilon_f$.

To compute the differential conductance, $dI/dV$
\cite{Mal09,Fig10a,Wol10}, measured in STS experiments
\cite{Schm09,Ayn10}, we define the spinor $\Psi^\dagger_{\bf
k}=(c^\dagger_{\bf k},f^\dagger_{\bf k})$ and the Green's function
matrix ${\hat G}({\bf k},\tau)=-\langle T_\tau \Psi_{\bf
k}(\tau)\Psi^\dagger_{\bf k}(0) \rangle$. With $t_c$ and $t_f$ being
the tunneling amplitudes into the $c$- and $f$-electron bands,
respectively, one has in the weak-tunneling limit \cite{Fig10a}
\begin{eqnarray}
\frac{dI({\bf r},\omega)}{dV} = - \frac{2 e}{\hbar} {\hat N}_t
\sum_{i,j=1}^2 \left[  {\hat t} \, {\rm Im} {\hat G}({\bf
r,r},\omega) \, {\hat t} \right]_{ij} \label{eq:dIdV}
\end{eqnarray}
where ${\hat t} =
\begin{pmatrix}
t_c& 0 \\
0 & t_f
\end{pmatrix}$, and $N_t$ is the STS
tip's density of states. To gain insight into the momentum resolved
electronic structure of URu$_2$Si$_2$, Schmidt {\it et al.}
\cite{Schm09} performed a quasi-particle interference (QPI) analysis
via the substitution of U by Th atoms. The measured QPI intensity,
$S({\bf q},\omega)$, is given by the Fourier transform of $dI/dV$
into ${\bf q}$-space, yielding
\begin{eqnarray}
S({\bf q},\omega) &\equiv& \frac{dI({\bf q},\omega)}{dV} = \frac{2
\pi e}{\hbar} N_t \sum_{i,j=1}^2 \left[{\hat t} {\hat N}({\bf
q},\omega)
{\hat t} \right]_{ij} \ ; \nonumber \\
{\hat N}({\bf q},\omega)&=&-\frac{1}{\pi} \ {\rm Im} \int \frac{d^2
k}{(2 \pi)^2} {\hat G}({\bf k},\omega) {\hat U} {\hat G}({\bf
k+q},\omega) \ . \ \  \label{eq:QPI}
\end{eqnarray}
${\hat U} =
\begin{pmatrix}
U_c& 0 \\
0 & U_f
\end{pmatrix}$ and $U_c$ and $U_f$ are the Th atoms' scattering potential in
the $c$- and $f$-electron bands, respectively.

\begin{figure}[h]
\includegraphics[width=8.cm]{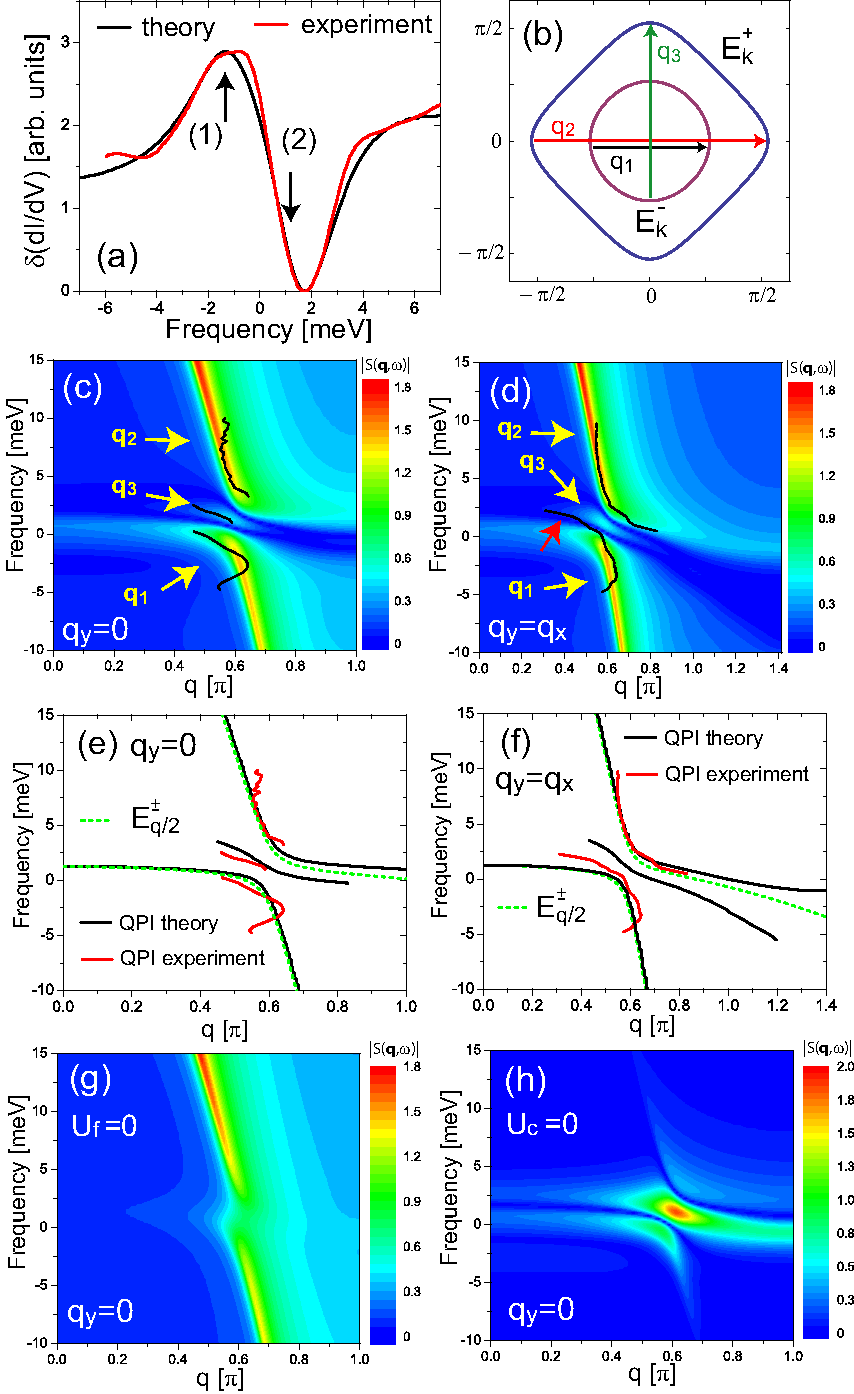}
\caption{(color online) (a) Experimental \cite{Schm09} and
theoretical $\delta(dI/dV)$ below the HOT. A background was
subtracted and the data were vertically scaled.  (b) Fermi surfaces
of $E_{\bf k}^{\pm}$. Contour plot of $|S({\bf q},\omega)|$ along
(c) $q_y=0$ and (d) $q_y=q_x$, together with the QPI dispersions of
Ref.~\cite{Schm09}. $|S({\bf q},\omega)|$ along $q_y=0$ for (e)
$U_f=0$ and (f) $U_c=0$. The theoretical results were obtained with
$t=45$ meV, $\mu=3.17t$, $s=0.06t$, $\varepsilon_f=-0.08t$,
$\chi_0=-0.04t$, $\chi_1=-0.3 \chi_0$, $\Gamma_{f,c}=0.03t$,
$t_f/t_c=0.0075$, $U_f/U_c=0.6$, yielding $J=2.95t$, $I=0.42t$,
$n_f=1.52$. } \label{fig:Fig1}
\end{figure}

We begin by discussing the STS results of Schmidt {\it et
al.}~\cite{Schm09} and present in Fig.~\ref{fig:Fig1}(a) the
experimentally measured change in $dI/dV$ below the HOT [$T=1.9$K
data of Fig.~3(b) in Ref.~\cite{Schm09}] together with a theoretical
fit, $\delta(dI/dV)=dI/dV(T<T_0)-dI/dV(T=T_0)$ (with $s=0$ at
$T=T_0$) obtained from Eq.(\ref{eq:dIdV}).
For the same parameter set as in Fig.~\ref{fig:Fig1}(a), we present
in Figs.~\ref{fig:Fig1}(c) and (e) [(d) and (f)] a contour plot of
the QPI intensity, $|S({\bf q},\omega)|$, and the maxima in $|S({\bf
q},\omega)|$ (i.e., the QPI dispersion), respectively, along $q_y=0$
[$q_y=q_x$]. Also shown are the experimental QPI dispersions (black
lines) of Figs.~5(c) and (d) in Ref.~\cite{Schm09}. The experimental
$dI/dV$ and QPI data were obtained on a U-terminated surface of a
1\% Th-doped sample. The very good quantitative agreement between
the theoretical and experimental $dI/dV$ and QPI dispersions
(arising from a unique set of parameters) strongly suggests that
their origin lies in the emergence of a coherent Kondo lattice below
the HOT, in agreement with the conclusions by Schmidt {\it et al.}
\cite{Schm09} (similar STS signatures of a coherent Kondo lattice
were recently also reported in YbRh$_2$Si$_2$ \cite{Ern10}).

The QPI pattern is determined by scattering of electrons both within
and between the two electronic bands, $E_{\bf k}^{\pm}$. Intraband
scattering [see Fig.~\ref{fig:Fig1}(b)] gives rise to the ${\bf
q}_1$ and ${\bf q}_2$ branches in $|S({\bf q},\omega)|$ shown in
Figs.~\ref{fig:Fig1}(c) and (d). The main contribution to these
branches arises from $2k_F$-scattering [Fig.~\ref{fig:Fig1}(b)],
such that their dispersion is approximately described by
$E^{\pm}_{\bf q/2}$ as shown in Figs.~\ref{fig:Fig1}(e) and (f).
Moreover, for $ -1 \mbox{ meV} \lesssim \omega \lesssim 1.5$ meV,
$E_{\bf k}^{\pm}$ both possess equal energy surfaces, giving rise to
interband scattering with wave-vector ${\bf q}_3$, and a
corresponding ${\bf q}_3$ branch in $|S({\bf q},\omega)|$ [see
Figs.~\ref{fig:Fig1}(c) and (d)]. The ${\bf q}_3$ branch has also
been seen experimentally along $q_y=0$, where the experimental and
theoretical QPI results are in very good agreement, but is absent
along $q_y=q_x$. The latter could arise from the smaller gap along
$q_y=q_x$ which might make it difficult to resolve the ${\bf q}_1$
and ${\bf q}_3$ branches. Additional support for this conclusion
comes the experimental $dI/dV$ data in Fig.~\ref{fig:Fig1}(a). Here,
the upper band edge of $E_{\bf k}^-$ leads to a sharp decrease in
$dI/dV$ [see arrow (2)] which occurs around $\omega \approx 1$ meV.
This energy is consistent with the extrapolation of the experimental
${\bf q}_1$ dispersion along $q_y=0$ [Fig.~\ref{fig:Fig1}(c)], where
the ${\bf q}_1$ and ${\bf q}_3$ branches are well resolved, but
inconsistent with the extrapolation along $q_y=q_x$. Note that the
peak in $dI/dV$ at $\omega = -2$ meV [see arrow (1) in
Fig.~\ref{fig:Fig1}(a)] arises from the van Hove singularity of the
$f$-electron band.

The experimental QPI pattern implies that the doped Th atoms scatter
electrons in both the $c$- and $f$-electron bands \cite{Schm09}. To
support this conclusion, we present in Fig.~\ref{fig:Fig1}(g) the
QPI pattern, $|S({\bf q},\omega)|$, for $U_f=0$. It is similar to
that of the unhybridized $c$-electron band [see
Fig.~\ref{fig:Fig3}(a)] since its dominant contribution arises from
scattering between states where the coherence factors of the
$c$-electrons are large. Conversely, for $U_c=0$,
[Fig.~\ref{fig:Fig1}(h)], $|S({\bf q},\omega)|$ is determined by
scattering between states with large $f$-electron weight. Both cases
are inconsistent with the experimental QPI results, whose
description requires $U_f/U_c \approx 0.6$, as shown in
Fig.~\ref{fig:Fig1}.

\begin{figure}[!h]
\includegraphics[width=8.cm]{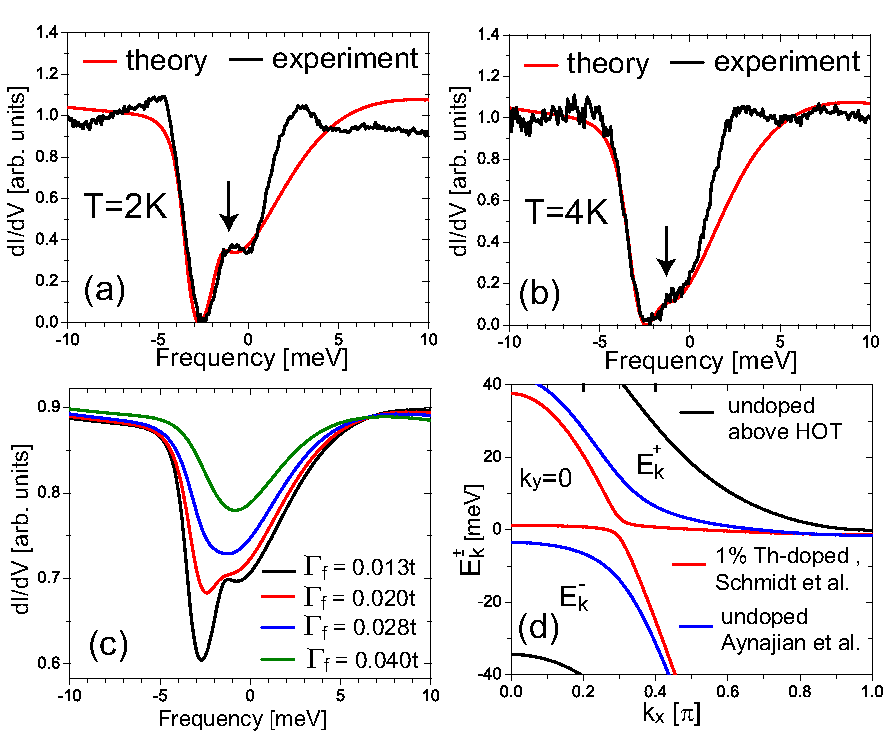}
\caption{(color online)  Theoretical fits to the $dI/dV$ data of
Ref.~\cite{Ayn10} at (a) $T=2$K and (b) $T=4$K. (c) Evolution of
$dI/dV$ with $\Gamma_f$. The theoretical results were obtained with
$J=3.69t$, $I=0.89t$, $n_f=1.59$, $\chi_1=-0.36 \chi_0$,
$t_f/t_c=0.0175$, and $\Gamma_c=0.02t$. At $T=2$K, $s=0.32t$,
$\varepsilon_f=-0.20t$, $\chi_0=-0.09t$. (d) $E_{\bf k}^\pm$
extracted from theoretical fits. } \label{fig:Fig2}
\end{figure}

\begin{figure}[!h]
\includegraphics[width=8.cm]{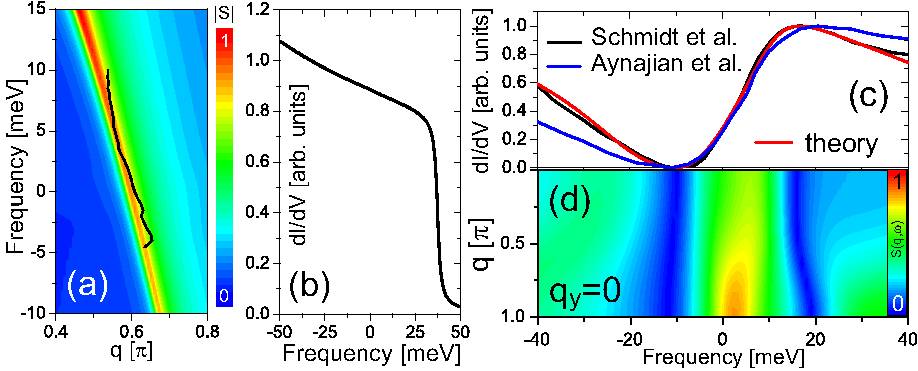}
\caption{(color online) (a) $|S({\bf q},\omega)|$ for $q_x=q_y$
together with the experimental QPI dispersion of Ref.~\cite{Schm09},
and (b) $dI/dV$, both for the same parameters as in
Fig.~\ref{fig:Fig1} but $s=0$. (c) Theoretical fit to the $19$K
$dI/dV$ data [Fig.5(c) in Ref.~\cite{Schm09}] with $s=1.0t$,
$\chi_0=-0.045t$, $\chi_1=0$, $\varepsilon_f=-0.032t$,
$\Gamma_c=0.77t$, $\Gamma_f=0.13t$, and (d) the resulting $|S({\bf
q},\omega)|$ along $q_y=0$. } \label{fig:Fig3}
\end{figure}

We next discuss the STS results by Aynajian {\it et al.}
\cite{Ayn10}, and present in Figs.~\ref{fig:Fig2}(a) and (b) the
experimental $dI/dV$ data obtained on a U-terminated surface of pure
URu$_2$Si$_2$ for $T=2$K and $4$K, respectively [Fig.~4(b) in
Ref.~\cite{Ayn10}], together with the theoretical results obtained
from Eq.(\ref{eq:dIdV}). The theoretical $dI/dV$ curves reproduce
the salient features of the experimental results: the asymmetry and
magnitude of the gap in $dI/dV$ as well as the peak at $\omega
\approx -0.8$ meV [see arrows in Figs.~\ref{fig:Fig2}(a) and (b)]
which arises from the van Hove singularity of the $f$-electron band
(a similar peak and gap magnitude were also reported for a
Si-terminated surface in Ref.\cite{Schm09}).
Both, the gap asymmetry and the peak, are characteristic signatures
of the Kondo lattice bandstructure, and thus suggest the existence
of a coherent Kondo lattice below the HOT. To gain further insight
into the microscopic origin of the CKL, we note that the changes in
$dI/dV$ between $T=2$K [Fig.~\ref{fig:Fig2}(a)] and $T=4$K
[Fig.~\ref{fig:Fig2}(b)] can be solely attributed to an increase in
the damping of the $f$-electron states from $\Gamma_f=0.013 t$ at
$T=2$K to $\Gamma_f=0.02 t$ at $T=4$K, and the resulting changes in
$s$, $\chi_0$ and $\varepsilon_f$ which are self-consistently
computed from Eq.(\ref{eq:SC2}) for fixed $J, I$ and $n_f$.
Increasing $\Gamma_f$ even further (while self-consistently
computing $s$, $\chi_0$ and $\varepsilon_f$) yields the evolution of
$dI/dV$ shown in Fig.~\ref{fig:Fig2}(c) which possesses the same
characteristic signatures as those observed by Aynajian {\it et al.}
\cite{Ayn10} with increasing temperature. In particular, the gap in
$dI/dV$ is filled in, its magnitude remains approximately constant
(until close to the HOT), and the center of the gap shifts to larger
energies with increasing temperature or $\Gamma_f$. Both $s$ and
$\chi_0$ decrease with increasing $\Gamma_f$ (not shown) since the
increased decoherence of the $f$-electron states necessarily
suppresses coherent Kondo screening and magnetic correlations. The
results in Figs.~\ref{fig:Fig2}(a)-(c) suggest that the formation of
a coherent Kondo lattice below the HOT is driven by a drastic
increase in the $f$-electron lifetime. In contrast, increasing
solely the hybridization, $s$, below the HOT leads to an evolution
of $dI/dV$ (not shown) that is inconsistent with the experimental
observations.

In Fig.\ref{fig:Fig2}(d), we present the bandstructure, $E_{\bf
k}^\pm$, obtained from the fits to the STS data by Schmidt {\it et
al.} for the 1\% Th-doped sample [Fig.~\ref{fig:Fig1}] and by
Aynajian {\it et al.} \cite{Ayn10} for pure URu$_2$Si$_2$
[Fig.\ref{fig:Fig2}(a)]. In the Th-doped sample, the hybridization
($s=0.06t$) is smaller while $\Gamma_{f,c}$ are larger than in the
undoped  compound ($s=0.32t$). These results are consistent with a
suppression of the HOT (and thus $s$) and an increased decoherence
by Th-doping, and thus suggest that both groups probe the same heavy
and light bands (while irrelevant for our study, we note that the
groups disagree on the identification of the cleaved surfaces).
Further support for this conclusion comes from the extracted
$f$-electron density, which is quite similar in both cases with
$n_f=1.59$ in the pure and $n_f=1.52$ in the Th-doped sample. While
these values deviate from the constraint $n_f=1$ usually enforced in
the Kondo limit \cite{largeN}, the good agreement between the
theoretical and experimental results suggest that corrections due to
valence fluctuations are small.

Finally, we discuss the $dI/dV$ Fano-lineshape observed above the
HOT \cite{Schm09,Ayn10} and its relation to the conduction band
observed in QPI \cite{Schm09}. To this end, we extend our analysis
of the STS results by Schmidt {\it et al.}\cite{Schm09} by assuming
a vanishing hybridization, $s=0$, for $T>T_0$. The resulting QPI
intensity $|S({\bf q},\omega)|$ shown in Fig.~\ref{fig:Fig3}(a)
reproduces well the experimental QPI dispersion (black line) of
Fig.~5(b) in Ref.~\cite{Schm09}. Due to the smallness of
$t_f/t_c=0.0075$, and since $s=0$, the contribution of the heavy
$f$-electron band to $|S({\bf q},\omega)|$ and $dI/dV$ is
negligible, thus explaining its absence in the experimental QPI data
above the HOT \cite{Schm09}. However, the resulting $dI/dV$
lineshape does not exhibit the characteristic Fano form, implying
that its origin resides in electronic bands not yet seen in QPI (the
sharp drop in $dI/dV$ at $\omega \approx 40$meV signifies the upper
band edge of $\varepsilon_{\bf k}$). To further investigate this
possibility, we present in Fig.~\ref{fig:Fig3}(c) a theoretical fit
using Eq.(\ref{eq:dIdV}) to the experimental $dI/dV$ data of
Ref.~\cite{Schm09} above the HOT on a Si-terminated surface in a
pure sample, which are similar to those of Ref.~\cite{Ayn10} on a
U-terminated surface. The resulting bands, $E_{\bf k}^\pm$, shown in
Fig.~\ref{fig:Fig2}(d), are not only significantly different from
the ones seen in QPI below the HOT, but also exhibit much larger
quasi-particle dampings, $\Gamma_{f,c}$, thus representing an {\it
incoherent Kondo lattice}. As a result, $|S({\bf q},\omega)|$, shown
in Fig.~\ref{fig:Fig3}(d), exhibits very little ${\bf q}$-structure
(for fixed $\omega$), thus explaining the difficulty in detecting
these bands in QPI. We therefore conclude that an explanation of the
STS data above and below the HOT requires multiple sets of $c$- and
$f$-electron bands.

We have shown that the STS results by Schmidt {\it et al.}
\cite{Schm09} and Aynajian {\it et al.} \cite{Ayn10} are consistent
with the emergence of a coherent Kondo lattice below the HOT in
URu$_2$Si$_2$. While it is not expected that the CKL is the primary
origin of the HOT \cite{Hew97}, it could be a result of the HOT. In
particular, one might speculate that the emergence of a yet unknown
order parameter at $T_0$ could lead to a significant decrease in the
quasi-particle decoherence, for example, through the gapping of
low-energy excitations, and thus induce the formation of a coherent
Kondo lattice, as described above. Clearly, further studies are
required to investigate this possibility.

We would like to thank P. Aynajian, J.C. Davis, and A. Yazdani for
stimulating discussions and comments, and for providing us with
their experimental data. This work is supported by the U.S.
Department of Energy under Award No. DE-FG02-05ER46225.

%\vspace{-0.5cm}

\end{document}